\documentclass[twocolumn,showpacs,preprintnumbers,amsmath,amssymb,floatfix,aps]{revtex4}

\usepackage{graphicx}
\usepackage{dcolumn}
\usepackage{bm}

\begin{document}

\preprint{APS/123-QED}

\title{Suppression of Tunneling in a Superconducting Persistent-Current Qubit}

\author{Rakesh P. Tiwari and D. Stroud}
\affiliation{%
Department of Physics, Ohio State University, Columbus, OH 43210\\
}%

\date{\today}

\begin{abstract}

We consider a superconducting persistent-current qubit consisting of a 
three-junction superconducting loop in an applied
magnetic field.  We show that by choosing the field, Josephson couplings, and offset charges
suitably, we can perfectly suppress the tunneling between the two oppositely directed states
of circulating current, leading to a vanishing of the splitting between the two qubit states.  This
suppression arises from interference between tunneling along different paths, and is analogous
to that predicted previously for magnetic particles with half integer spin.

\end{abstract}

\maketitle
Surprising physical effects can
be produced by quantum-mechanical interference between particles moving from one site to another
along different paths\cite{geometricphases,berry}.  Examples include the Aharonov-Bohm effect\cite{ab}
and the Aharonov-Casher effect\cite{ac,ra}.  Many authors have considered this Aharonov-Casher effect for Josephson-junction arrays and devices\cite{bjvanwees,esimanek,ivanov,friedman}. 
Starting from this topological effect Loss et al.\cite{ldvg} predicted suppression of tunneling due to interference of different tunneling paths for magnetic particles with half integer spin, and also oscillations in tunnel splitting with applied magnetic field. These oscillations were confirmed experimentally by Wernsdorfer and Sessoli \cite{ws}.   

Effects related to those analyzed in Ref.\ \cite{ldvg} have also been studied
in systems of Josephson junctions\cite{ivanov,oetal,blatter}.  For example, a
three-junction loop has been studied 
as a possible phase qubit\cite{oetal,blatter}.
In its original design\cite{oetal}, the junctions were deliberately made asymmetric to avoid interference of different tunneling paths, and to protect the qubit from random charge fluctuations.

In this note, we analyze the same three-junction loop but in a different regime, namely, one in 
which the asymmetry is chosen to observe and control the interference of tunneling paths.   We show that for this chosen asymmetry,
the interference is easily detected because the tunnel splitting 
{\it vanishes perfectly} for certain special induced gate charges.   The suppression of the tunneling splitting
is closely analogous to that discussed in Ref.\ \cite{ldvg}.  Although this qubit is potentially
subject to random charge noise, it may be possible to minimize this noise
by appropriately tuning the gate voltage\cite{mooij}.  

\begin{figure}[ht]
\begin{center}
\includegraphics[scale=0.6,angle=0]{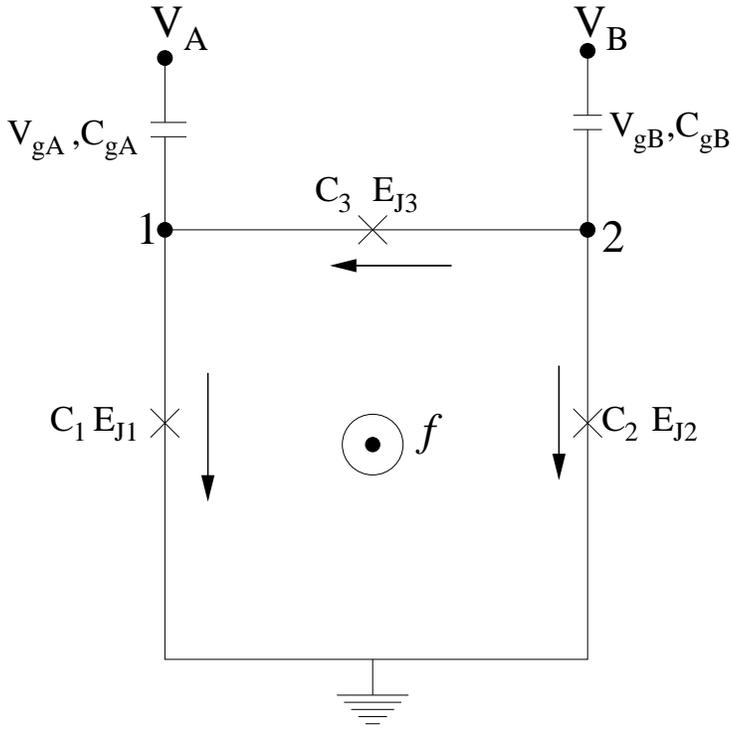}
\caption{Schematic of the circuit for the three-junction qubit, after Ref.\
\cite{oetal}.  There are two superconducting islands, denoted $1$ and $2$,
whose voltages are $V_1$ and $V_2$.  The three junctions in the circuit are indicated
by crosses; the i$^{th}$ junction has capacitance $C_i$ and Josephson coupling energy
$E_{Ji}$. An external flux $\Phi = f\Phi_0$ passes through the circuit. The superconducting islands
$1$ and $2$ are also connected to applied voltages $V_A$ and $V_B$ through capacitors
$C_{gA}$ and $C_{gB}$; the voltage across these capacitors is $V_{gA} = V_A - V_1$ and $V_{gB} = V_B - V_2$.}\label{fig:1}
\end{center}
\end{figure}

The circuit diagram for the three-junction qubit of Ref.\ \cite{oetal} is shown in Fig.\ \ref{fig:1}.  
The i$^{th}$ junction (i = 1, 2, 3) has
capacitance $C_i$ and Josephson coupling energy $E_{Ji}$.   An external
flux $\Phi = f\Phi_0$, where $\Phi_0 = h/(2e)$, is applied through the three-junction loop, which is assumed 
to have negligible self-inductance.  The properties of
the qubit can be manipulated by controlling $\Phi$, and also 
the two external voltages, $V_A$ and $V_B$, which
are applied to the circuit through two gate capacitors $C_{gA}$ and $C_{gB}$.  The voltages across
these capacitors are $V_{gA} = V_A - V_1$ and $V_{gB} = V_B - V_2$.  

We assume, following Ref.\ \cite{oetal}, that junctions 1 and 2 have equal Josephson energies
$E_J$ and equal capacitances $C$, while junction 3 has Josephson energy $\alpha E_J$
and capacitance $\alpha C$.  We also assume that $C_{g,A} = C_{g,B} = \gamma C$.   Because of the fluxoid quantization, the three gauge-invariant phase
differences $\phi_1$, $\phi_2$, and $\phi_3$ satisfy the constraint $\phi_1 - \phi_2 + \phi_3 = -2\pi f$.
We choose $\phi_1$ and $\phi_2$ to be the independent dynamical variables, and write $\phi_3 = 2\pi f + \phi_1 - \phi_2$.

With these choices, the circuit Lagrangian ${\cal L}$ can be written as
$\mathcal{L}=\mathcal{T}-\mathcal{U}-(\frac{\Phi_{0}}{2\pi}) \overline{\dot{\phi}}^{T}\cdot {\bf C}_{g}\cdot\overline{V}_{g}=\mathcal{L}_{0}+\mathcal{L}_{WZ}$,
where $\mathcal{L}_0 = \mathcal{T}-\mathcal{U}$ and $\mathcal{L}_{WZ}$ is the remaining term, 
which we call the Wess-Zumino term.
The kinetic energy term $\mathcal{T}$ represents the electrical energy stored in all the capacitors of the
system; it can be written as
$\mathcal{T}=\frac{1}{2}(\frac{\Phi_{0}}{2\pi})^{2} \overline{\dot{\phi}}^{T}\cdot{\bf C}\cdot\overline{\dot{\phi}}$
The Josephson coupling energy 
$\mathcal{U}=E_{J}\left[2+\alpha-\cos\phi_{1}-\cos\phi_{2}-\alpha \cos(2\pi\textsl{f}+\phi_{1}-\phi_{2})\right]$.
Finally, $\mathcal{L}_{WZ}=(\frac{\Phi_{0}}{2\pi}) \overline{\dot{\phi}}^{T}\cdot {\bf C}_{g}\cdot\overline{V}_{g}$.

In the above expression for ${\cal L}$, the quantity {\bf C} represents the capacitance matrix
for the junctions, with matrix elements $C_{11} = C_{22} = C(1 + \alpha + \gamma)$, $C_{12} = C_{21} =
-\alpha C$.
Also, $\overline{\dot{\phi}}$ represents the ($1 \times 2$)
column vector with elements $\dot{\phi}_1$, $\dot{\phi}_2$; $\overline{V}_{g}$ is the $1 \times 2$ column vector with elements $V_A$ and $V_B$; and the gate capacitance matrix ${\bf C}_{g}$ is a $2\times2$ diagonal matrix with both diagonal elements equal to $\gamma C$.
All the above formalism is identical to that in \cite{oetal}.   However, we consider a different set of parameters: $\alpha > 1$ (rather than $\alpha < 1$) and
$f \sim 1/2$.  This difference has a qualitative effect on the qubit behaviour.

\begin{figure}[ht]
\begin{center}
\includegraphics[scale=0.6,angle=0]{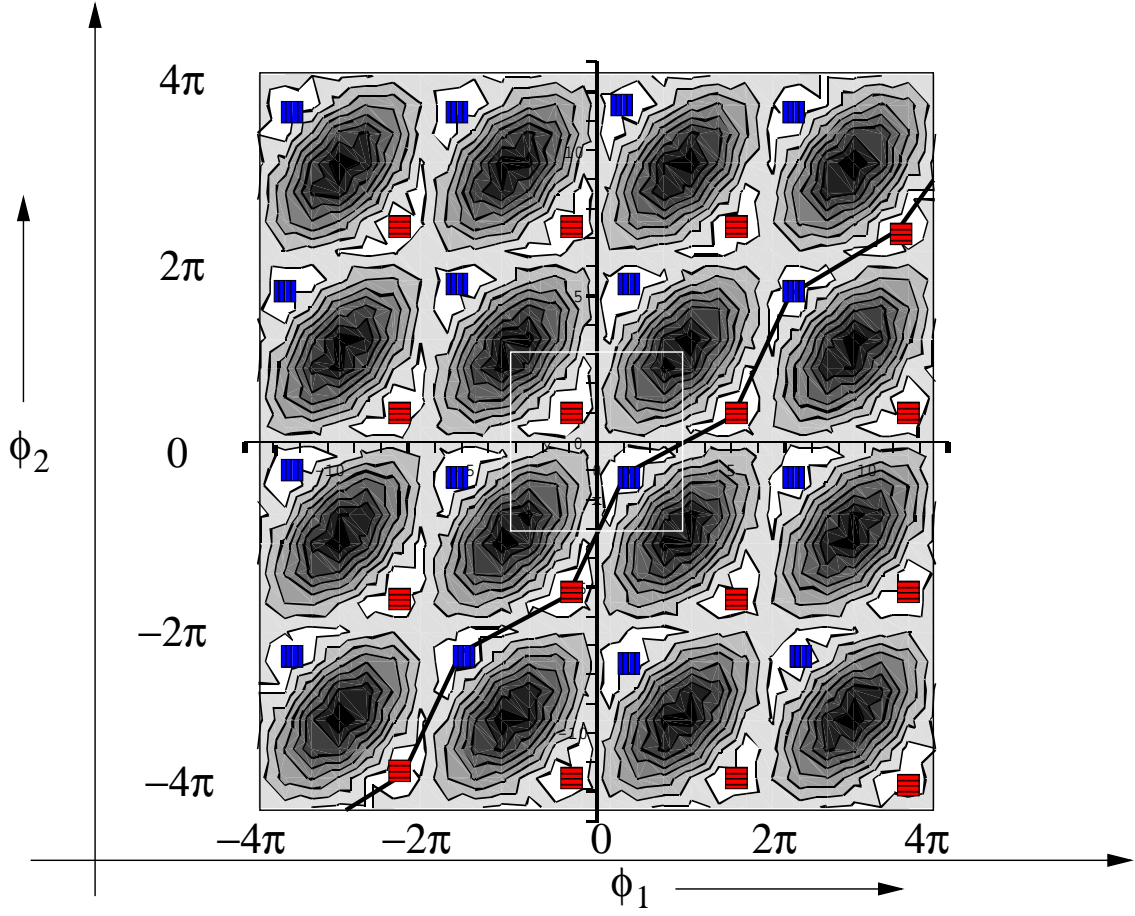}
\caption{(Color online) Contour plot of the potential $\mathcal{U}(\phi_{1},\phi_{2})$ for the special case
$\alpha = 1.3$ (see text).   The horizontal and
vertical axes represent $\phi_{1}$ and $\phi_{2}$.  Darker shading means larger value of $\mathcal{U}$.  For this choice of $\alpha$=1.3, the state denoted by a box with vertical lines
in the lower right of the white square can tunnel to another state only along the heavy line; for other directions, the tunneling barrier is much higher.  The white square represents the unit cell used in the text.} \label{fig:2}
\end{center}
\end{figure}

Fig.\ \ref{fig:2} shows a contour plot of the potential energy ${\cal U}(\phi_1, \phi_2) $ for $\alpha = 1.3$ and $f = 1/2$, represented in the repeated cell scheme.  The origin ($\phi_1 = 0$, $\phi_2 = 0$) is shown
at the center of the plot.   The potential energy is periodic in both $\phi_1$ and
$\phi_2$ with period $2\pi$.   With our choice of parameters,
this potential energy has two inequivalent states of the same minimum energy, 
indicated by boxes with horizontal and vertical lines within each unit cell. 
If we choose the unit cell to be $-\pi < \phi_1 < \pi$, $-\pi < \phi_2 < \pi$, then the two inequivalent
minima are in the upper left and lower right quadrants of the cell.     
These two minima occur at ($\phi^{\ast} + 2m\pi,-\phi^{\ast} + 2n\pi$) and $(-\phi^{\ast} + 2m\pi, \phi^{\ast} + 2n\pi)$  where $m$ and $n$ are positive or negative integers and $\phi^{\ast}=\cos^{-1}\frac{1}{2\alpha}$. 
These two states are degenerate, but {\em inequivalent}.  Physically, they correspond to states with clockwise and counterclockwise loop currents.  
When $\alpha > 1$,
the lowest-barrier tunneling paths between these states are different than in the $\alpha < 1$ case considered
in Ref.\ \cite{oetal}.   

To see this, suppose we start from the state ($\phi^{\ast},-\phi^{\ast}$),
and suppose that this state represents a clockwise-circulating loop current,
corresponding to boxes containing vertical lines in
the lower right-hand corner of the central unit cell.
There are three plausible tunneling directions to
reach a neighboring state with counterclockwise-circulating currents,
leading to states at ($2\pi - \phi^\ast,
\phi^\ast$), ($-\phi^\ast$, $-2\pi + \phi^\ast$), and $(-\phi^\ast$, $\phi^\ast$).
If $\alpha > 1$, one can show numerically that 
the potential barrier is smaller for the two located at ($2\pi-\phi^{\ast},\phi^{\ast}$) and ($-\phi^{\ast},-(2\pi-\phi^{\ast}$)) than that for tunneling to ($-\phi^{\ast}$,$\phi^{\ast}$).
By contrast, if $\alpha < 1$, the tunneling barrier is smaller for the third path than for the
other two.   Because there are {\it two} possible lowest-barrier paths when $\alpha > 1$,
there is an interference effect in this case which is absent when $\alpha < 1$.  Furthermore, the difference in barrier heights (between
the two equal-barrier tunneling paths and the third, higher-barrier path) increases with increasing $\alpha$,
provided $\alpha > 1$.  Thus,
we can easily choose $\alpha$ so that the system tunnels only through these
barriers.  This tunneling corresponds to the paths near the
heavy line in Fig.\ \ref{fig:2}.   
We will show that, for $\alpha > 1$, there exist certain values $Q_A$ and $Q_B$
of the stored charge, for which the tunneling along these two equal barrier
paths {\it exactly cancels out}.


In the absence of tunneling, the system has two degenerate minimum-energy
quantum states when $f = 1/2$, one with counterclockwise ($\bigodot$) and the other
with clockwise ($\bigotimes$) current.  In the presence of tunneling, these two states
are connected by a tunneling matrix element $w$, which breaks the degeneracy.


The transition amplitude ($\mathcal{P}$) from a state $\bigotimes$ to
state $\bigodot$ can be calculated using the imaginary time coherent state
path integral method.  Symbolically, at temperature $T = 0$, 
$\mathcal{P}=\int_{\phi_1(0),\phi_2(0)}^{\phi_1(\infty),\phi_2(\infty)}{D\Omega}e^{-\frac{1}{\hbar}\mathcal{S}_{\phi_{1}\phi_{2}}}$,
where $D\Omega$ represents an integral over all paths in imaginary time
starting from the clockwise state at $(\phi_1(0),\phi_2(0)) = (\phi^\ast,-\phi^\ast)$ at $\tau = 0$ 
and ending at the counterclockwise state at $(\phi_1(\infty),\phi_2(\infty) = (2\pi - \phi^\ast, \phi^\ast)$ or
$(-\phi^\ast, 2\pi-\phi^\ast)$ at $\tau = \infty$. 
$\mathcal{S}_{\phi_{1}\phi_{2}}$ represents the action calculated along each of the paths.
In turn, $\mathcal{S}_{\phi_{1}\phi_{2}}=
\int_{\phi_1(0),\phi_2(0)}^{\phi_1(\infty),\phi_2(\infty)}d\tau({\mathcal{L}_{0}}+\mathcal{L}_{wz})$, 
where the integral is over imaginary times $\tau$ (such that $t = i\tau$), and $\mathcal{L}_0 + \mathcal{L}_{wz}$
is the Lagrangian but with each time $t$ replaced by $i\tau$.  At $T = 0$, the integrals start at $\tau = 0$ and
run to $\tau = \infty$. 

The key point is that, for $\alpha > 1$, there are two classes of paths going from the point $\bigotimes$ to
the point $\bigodot$ in phase space.  One of these is generally in the ``northeast'' (NE) direction and the
other in the ``southwest'' (SW) direction; the paths run in generally opposite directions in the vicinity of the heavy
black line in Fig.\ 2.  The two end-points of the paths in the NE direction are $\bigotimes = (\phi^\ast, -\phi^\ast)$, and $\bigodot = (2\pi - \phi^\ast,\phi^\ast)$, while for those in the SW direction they are $\bigotimes = (\phi^\ast, -\phi^\ast)$ and $\bigodot = (-\phi^\ast, -(2\pi - \phi^\ast))$.  Let us consider one particular path in the
NE direction, and denote it by $(\phi_1(\tau), \phi_2(\tau))$.
This path runs from $(\phi^\ast, -\phi^\ast)$ to $(2\pi -\phi^\ast, \phi^\ast)$.  Then the
path $-\phi_2(\tau), -\phi_1(\tau))$ also starts from $(\phi^\ast,-\phi^\ast)$ but runs generally in the
SW direction to $(-\phi^\ast, -(2\pi-\phi^\ast))$.  Thus, for every path in the NE direction, we can define a
corresponding path in the SW direction by this procedure.

We now show that the contributions of these two paths to the path integral exactly cancel out for special values
of $Q_A$ and $Q_B$.  We first consider the contributions of ${\cal U}$ and ${\cal T}$ to the path integral.
At any point along a NE path, the potential energy ${\cal U}(\phi_1, \phi_2)$, is given above.  Along any point along the corresponding SW path, the corresponding potential energy is given by
${\cal U}(-\phi_2, -\phi_1)$, which from eq.\ (3) equals ${\cal U}(\phi_1, \phi_2)$.  Thus, the contribution
of ${\cal U}$ to $S$ is exactly the same for corresponding paths in the NE and SW directions.  Similarly, the contribution of ${\cal T}$ to $S$ is the same for 
corresponding paths in the NE and SW directions (because ${\cal T}$ is quadratic in the derivatives
$\dot{\phi}_1$ and $\dot{\phi}_2$, and because the diagonal elements of ${\cal C}$ are equal). 
Since $S$ appears in the exponential, the exponential $\exp(-S/\hbar)$ terms give the same multiplicative contribution to $\mathcal{P}$
for each of the two paths.


For $\mathcal{L}_{WZ}$, we have
$\mathcal{L}_{WZ}=-i\left(\frac{\Phi_{0}}{2\pi}\right)\gamma C (V_{A} \dot{\phi_{1}}+ V_{B} \dot{\phi_{2}})$,
where $\dot{\phi_{1}}$ and $\dot{\phi_{2}}$ are derivatives with respect to $\tau$.
Since $\mathcal{L}_{WZ}$ is a total time derivative, its contribution to 
$\mathcal{S}_{\phi_{1}\phi_{2}}$ depends only on the initial and final values
$i$ and $f$ of the phases $(\phi_1$ and $\phi_2)$.  From eq.\ (4), this contribution is
$-i\frac{\Phi_{0}}{2\pi}\int_{i}^{f}d\tau(\gamma C (V_{A} \dot{\phi_{1}}+ V_{B} \dot{\phi_{2}}))=-i(\frac{\Phi_{0}}{2\pi})\gamma C[V_{A}(\phi_{1}(f)-\phi_{1}(i))+V_{B}(\phi_{2}(f)-\phi_{2}(i))]$.
Thus, for any path taking the state ($\phi^{\ast}$,$-\phi^{\ast}$) in the NE direction to ($2\pi-\phi^{\ast}$,$\phi^{\ast}$), $\mathcal{L}_{WZ}$ gives a contribution to $S$ equal to 
$\mathcal{S}_{WZ}^{NE}=-i\left(\frac{\Phi_{0}}{2\pi}\right)\gamma C\left[2V_{A}(\pi-\phi^{\ast})+2V_{B}
\phi^{\ast}\right]$.
Similarly, a path taking the state ($\phi^{\ast}$,$-\phi^{\ast}$) in the SW direction to ($-\phi^{\ast}$,$-(2\pi-\phi^{\ast})$) gives a contribution  
$\mathcal{S}_{WZ}^{SW}=i\left(\frac{\Phi_{0}}{2\pi}\right)\gamma C\left[2V_{A}\phi^{\ast}+2V_{B}(\pi-
\phi^{\ast})\right]$.  

We now write $\frac{\Phi_{0}}{2\pi}\gamma C V_{A} = \hbar\frac{Q_{A}}{2e}$ and $\frac{\Phi_{0}}{2\pi}\gamma C V_{B} = \hbar\frac{Q_{B}}{2e}$, where $Q_{A}$ and $Q_{B}$ represent the charge stored on the gate capacitors, and $e$ represents the electronic charge.
Then the sum of the contributions of $\mathcal{S}_{WZ}^{NE}$ and $\mathcal{S}_{WZ}^{SW}$ to $\mathcal{P}$ is
$\mathcal{P}_{WZ}=\mathcal{P}_{0}(e^{\frac{i}{e}[Q_{A}(\pi-\phi^{\ast})+Q_{B}\phi^{\ast}]}+e^{-\frac{i}{e}[Q_{A}\phi^{\ast}+Q_{B}(\pi-\phi^{\ast})]})$,
where ${\cal P}_0$ is a constant term which is the same for the two paths.  It is always
possible to define a number {\it n} such that $Q_B = ne - Q_A$.  In terms of $n$, we can rewrite $\mathcal{P}_{WZ}$ as
$\mathcal{P}_{WZ} = \mathcal{P}_0\exp\left[\frac{i}{e}Q_A(\pi - 2\phi^*) + in\phi^*\right]
\left[1 + e^{-in\pi}\right]$.  Hence, 
$\mathcal{P}_{WZ}$ {\em vanishes} whenever $Q_A + Q_B = ne$, where $n$ is an odd integer.  Since
the contributions of $\mathcal{U}$ and $\mathcal{T}$ to $S$ are the same for corresponding paths in NE and SW directions, the total $\mathcal{P}$ still vanishes when this condition is satisfied, even including the 
contributions of
$\mathcal{U}$ and $\mathcal{T}$ to $S$.  Our calculation is analogous to that of Loss {\it et al.}\cite{ldvg} for a magnetic tunneling problem.


Thus, the paths taking the state ($\phi^{\ast}$,$-\phi^{\ast}$) to ($2\pi-\phi^{\ast}$,$\phi^{\ast}$) and to ($-\phi^{\ast}$,$-(2\pi-\phi^{\ast})$) interfere 
{\it completely destructively} whenever the stored charges on the gate capacitors sum to an odd multiple of $e$. This destructive interference is not restricted to straight line paths, because, as we have shown, for any general path in the NE direction, there exists a path in the SW direction which interferes destructively with it. 
It can be shown that similar destructive interference occurs for higher order paths, such as the next order paths which take the state ($\phi^{\ast}$,$-\phi^{\ast}$) to ($4\pi-\phi^{\ast}$, $2\pi+\phi^{\ast}$) and ($-(2\pi+\phi^{\ast})$, $-(4\pi-\phi^{\ast})$). 
This implies that the two persistent current(clockwise and counterclockwise) states remain degenerate for these
values of the stored charges, provided that $f = 1/2$.



The same cancellation can be demonstrated using the tight-binding formulation of
Ref.\ \cite{oetal}.   The classical Hamiltonian ${\cal H}$ corresponding to the above Lagrangian  is 
${\cal H} = \overline{p}^T\cdot\overline{\dot{\phi}} - {\cal L}$, where the canonical momenta 
$p_i =\frac{\partial{\cal L}}{\partial \dot{\phi}_i}$, i = 1, 2.
  This procedure gives\cite{oetal}
${\cal H} = \frac{1}{2}\left(\frac{2\pi}{\Phi_0}\right)^2\overline{P}^T\cdot{\bf C}^{-1}\cdot\overline{P}
+ {\cal U}(\phi_1, \phi_2)$,
where $\overline{P} = \overline{p} + \frac{\Phi_0}{2\pi}{\bf C}_g\cdot\overline{V}_g$,
and ${\cal U}$ is given above.

The energy eigenstates of ${\cal H}$ satisfy the time-independent Schr\"{o}dinger equation
${\cal H}\psi(\phi_1, \phi_2) = E\psi(\phi_1, \phi_2)$,
where $p_i = -i\hbar(\partial/\partial\phi_i)$ (i = 1, 2).  The boundary
conditions, obtained from the requirement that the wave function be single-valued, are
$\psi(\phi_1 + 2m\pi, \phi_2 + 2n\pi) = \psi(\phi_1, \phi_2)$,
where $m$ and $n$ are integers. 
  
The charge periodicity discussed above is due to the relation between the components of
$\overline{P}$ and $\overline{p}$.  Specifically, using the fact that the matrix
$\overline{C}_g$ is diagonal, we write $P_1 = p_1 + \frac{\Phi_0}{2\pi}C_{g,A}V_{g,A}= -i\hbar\left(\frac{\partial}{\partial\phi_1} + i\frac{Q_A}{2e}\right)$,
$P_2 = p_2 + \frac{\Phi_0}{2\pi}C_{g,B}V_{g,B}= -i\hbar\left(\frac{\partial}{\partial\phi_2} + i\frac{Q_B}{2e}\right)$.  on using the operator forms of 
$p_1$ and $p_2$ and using $Q_A = C_{g,A}V_{g,A}$ and 
$Q_B = C_{g,B}V_{g,B}$.  We now define $\chi(\phi_1, \phi_2) = \exp(i{\bf k}\cdot\overline{\phi})
\psi(\phi_1,\phi_2)$, where the vector ${\overline\phi} = \phi_1\hat{\phi}_1 + \phi_2\hat{\phi}_2$, ${\bf k} = \frac{Q_A}{2e}\hat{\phi}_1 + \frac{Q_B}{2e}\hat{\phi}_2$, and
$\hat{\phi}_1$ and $\hat{\phi}_2$ are unit vectors in the $\phi_1$ and $\phi_2$ directions
in Fig.\ 2.  Then the boundary conditions
on $\psi$ imply that $\chi(\phi_1, \phi_2)$ is a Bloch function, i.\ e., that $\chi(\phi_1 + 2\pi m,\phi_2 + 2\pi n) = \exp(i{\bf k}\cdot{\bf R_{mn}})\chi(\phi_1, \phi_2)$, where ${\bf R} = 2\pi m\hat{\phi}_1 + 2\pi n\hat{\phi}_2$ is a lattice vector. Also, in terms of $\chi$, the
two-variable Schr\"{o}dinger equation takes the form $({\cal T} + {\cal U})\chi_{\bf k} = E({\bf k})\chi_{\bf k}$.
Since $\chi_{\bf k}$ is a Bloch function and
${\cal U}$ is periodic, the eigenvalue $E({\bf k})$ is periodic in ${\bf k}$ and hence, in the charges $Q_A$ and $Q_B$.

This Schr\"{o}dinger equation can be solved within a tight-binding approximation\cite{oetal} to calculate
the tunnel splitting, which we previously calculated using a path integral approach.  Let us consider two localized ``atomic'' orbitals, $u(\phi_1, \phi_2)$ and $v(\phi_1, \phi_2)$, which represent the ground state wavefunctions in each of the two minima of the potential ${\cal U}(\phi_1, \phi_2)$ within the central unit cell.   Then, 
at Bloch vector ${\bf k}$ ,the tight binding wavefunction 
$\chi_{\bf k}(\phi_1, \phi_2)=c_{{\bf k},u}u(\phi_1, \phi_2)+c_{{\bf k},v}v(\phi_1, \phi_2)$,	
where $c_{{\bf k},u}$ and $c_{{\bf k},v}$ are defined by
$[H_{uu}({\bf k})-E({\bf k})]c_{{\bf k},u}+H_{uv}({\bf k})c_{{\bf k},v} = 0$,
$H_{vu}({\bf k})c_{{\bf k},u}+[H_{vv}({\bf k})-E({\bf k})]c_{{\bf k},v} = 0$.
In actuality, $H_{uu}$ and $H_{vv}$ are independent of ${\bf k}$.  

When the applied field is such that $f = 1/2$, $u$ and $v$ are exactly degenerate, with energy
which we denote $\epsilon_{0}$.  In this case, $H_{uu}=H_{vv}=\epsilon_{0}$.  To obtain the
other two elements, we denote by $t_{1}$ the tunneling matrix element between these two minima in the same unit cell, and $t_{2}$ between the state in the ``southeast'' corner of that cell and either of the
two adjacent minima lying along the heavy line in Fig.\ 2.  Then 
$H_{uv}=H^{\ast}_{vu}=-t_{1}-t_{2}\left[e^{i{\bf k}\cdot{\bf R}_1}+e^{i{\bf k}\cdot{\bf R}_2}\right]$,
where ${\bf R}_1 = 2\pi\hat{\phi_{1}}$ and ${\bf R}_2 = -2\pi\hat{\phi_{2}}$ are the Bravais lattice vectors from the central unit cell (denoted
by a white square in Fig.\ 2) to the two adjacent cells along the heavy line in Fig.\ 2.

An estimate of the $t_{i}$'s can be obtained using the WKB method, by calculating the action $S_{i}$ 
between the two minima and writing $t_{i}\approx(\hbar\omega_{i}/2\pi)e^{-S_{i}/\hbar}$.  Here $\omega_{i}$ is the attempt frequency for the phase ``particle'' to escape from the potential well. 
Following the approach of Ref. \cite{oetal}, we find that for $\alpha=1.3$ and $E_{J}/E_{C}\sim 100$, where $E_{C}=e^{2}/(2C)$, the ratio $t_{1}/t_{2}\sim 10^{-4}$.  In
fact, provided $E_J/E_C \ll 1$, we can always choose an $\alpha > 1$ such that the effect of $t_{1}$ is
very small.

Neglecting $t_1$ and using the above values of ${\bf k}$, ${\bf R}_1$ and ${\bf R}_2$, we obtain  
$H_{uv}= -2t_2\exp[i\pi(Q_A-Q_B)/(2e)]\cos[\pi(Q_A+Q_B)/(2e)]$.  
The eigenvalues of $H$ are then $E=\epsilon_{0}\mp\left|H_{uv}\right|$, with corresponding
normalized eigenvectors $(u \pm v)/\sqrt{2}$.    The result for the eigenvalues shows 
that, when the offset charges satisfy $Q_A + Q_B = ne$, with $n$ an odd integer, the levels become degenerate.  This is exactly the result we found by our path interference analysis.  Note
that the energy splitting depends only on the {\em sum} $Q_A + Q_B$, not on the difference
$Q_A - Q_B$.


As $f$ deviates slightly from 1/2, the potential \textit{U} changes such that the two minimum states of the two wells in a unit cell
become unequal in energy, and the barrier heights also change. 
If we define the zero of energy as the average of the two lowest energy states at $f$=1/2, the elements
of $H$ become $H_{uu} = -H_{vv} = F$; $H_{uv} = H_{vu}^* = -t$.  
Here $F \sim (\partial H_{uu}/\partial f)(f - 1/2)$ is the
change in the diagonal matrix element of $H$ with a small change in flux.  Also, if we write
$Q_A + Q_B = (2n+1)e + \delta Q$, and we assume $|\delta Q/e| \ll 1$, we find $t \sim -2t_2\exp[i\pi(Q_A-Q_B)/(2e)](-1)^n\pi\delta Q/(2e)$.  
Thus, to first order in $\delta f$ and $\delta Q$, $F$ and $t$ are controlled by {\it two} different parameters: $f - 1/2$ and $\delta Q/e$. The corresponding eigenvalues of $H$ are $E_{\mp} = \mp\sqrt{F^{2}+|t|^{2}}$, and 
depend on $f-1/2$ (through F) and $Q_A + Q_B$ (through t), but not on $Q_A - Q_B$.  (The eigenvectors do
depend on $Q_A - Q_B$.)  By manipulating
these two control parameters independently, one could, in principle, adjust the splitting of this two-level
system.  

\begin{figure}[ht]
\begin{center}
\includegraphics[scale=0.6,angle=0]{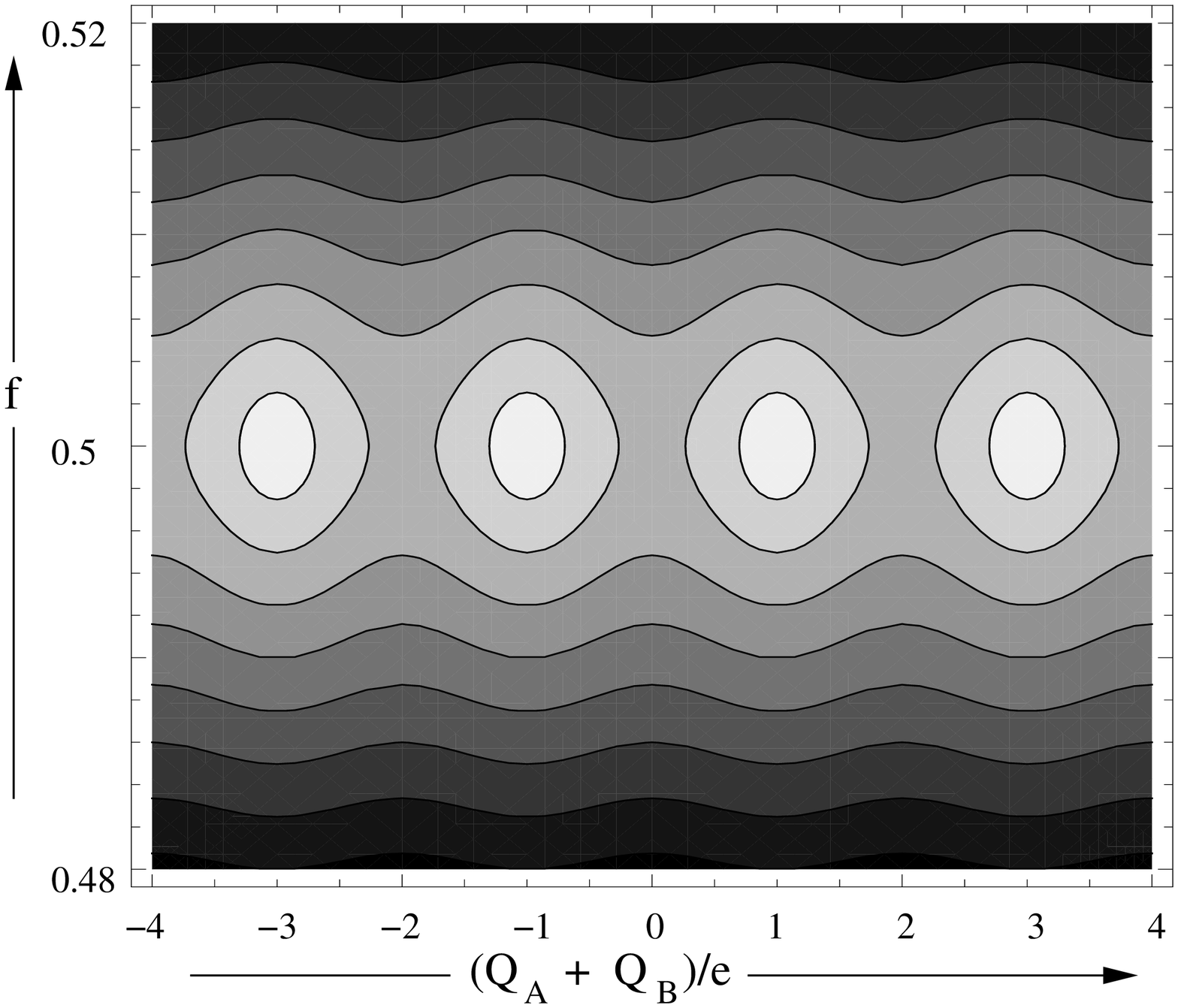}
\caption{Contour plot of the splitting $2\sqrt{F^{2}+|t|^{2}}$. The horizontal and
vertical axes represent $(Q_A + Q_B)/e$ and $f$.  Darker shading means larger values of the splitting.
Except for the value of $\alpha$, we 
have used the same parameters as in Ref.\ \cite{oetal} (see text).  
The splitting vanishes when $(Q_A + Q_B)/e$ is an odd integer and $f=1/2$.}
\label{fig:3}
\end{center}
\end{figure}

Fig.\ \ref{fig:3} shows a contour plot of the quantity $2\sqrt{F^{2}+|t|^{2}}$, which represents the energy difference
between the two lowest eigenvalues of $H$, as a function of the quantities $f$ and $(Q_A + Q_B)/e$.  In constructing
this plot, we assume the following parameters: 
$E_J/E_C = 80$, $\gamma = 0.02$, $\alpha = 1.3$, and an attempt frequency $\hbar\omega/(2\pi) = 0.193 E_J$.  Except for $\alpha$, all these
quantities have the same values as in Ref.\ \cite{oetal}.  There are eight contour curves visible, equally spaced
between $0$ and a maximum value of $0.3266 E_J$.  These are calculated using the above parameters and the
relation $[\partial H_{uu}/\partial f]_{f = 1/2} = 4\pi\alpha E_J\sin(2\pi f + 2\phi^*)$.  The splitting vanishes
when $(Q_A + Q_B)/e$ is an odd integer and $f = 1/2$.  

The parameters $F$ and $|t|$ should be controllable experimentally.  $F$ can be finely adjusted by changing
$f$, the magnetic flux through the loop. For the parameters of Ref.\ \cite{oetal}, $|t|$ should also be controllable. 
Taking $E_J = 800\mu eV$, we have $E_C = 10 \mu eV$, corresponding to a junction capacitance $C = 10 fF$, and hence
gate capacitances $C_g = 0.16 fF$.  With this value for $C_g$, $(Q_A + Q_B)/e = 1$ corresponds to 
$V_A + V_B = C_g(Q_A + Q_B) = 1mV$, a
value which should be tunable to a small tolerance.    In the different
regime of small junctions ($E_J \sim E_C$), the periodicity of energies with offset charges has been observed,
e.\ g., in Ref.\ \cite{mooij}.  In that work, a computer-controlled method was used to accurately compensate
for the random offset charges.  For operation of the present system as a qubit in the regime with $\alpha > 1$, one 
would need a temperature $T$ low enough to avoid creation of single-electron
excitations, or exciting the system above its two lowest levels.  Using the parameters of Ref.\ \cite{oetal}, 
this would be $k_BT \ll 0.2 E_J \sim 2 K$.  A temperature of 0.2-0.4 K should be sufficient, and
is readily attainable with current cryogenics.



To summarize, we have demonstrated that the three-junction persistent-current qubit can be placed in a
regime such that the states are determined by the interference of tunneling paths.  For certain values
of the offset charges, this interference is perfectly destructive, leading to a vanishing of the tunnel
splitting between the two states of the qubit for appropriate values of the gate charges and the
applied magnetic field.  This effect should be observable experimentally, as long as the
sum of the offset charges can be controlled experimentally, as we have briefly discussed above.  It would
certainly be of interest to observe the cancellation suggested here.  



This work was supported by NSF through grant DMR04-13395.

\newpage


\newpage

\begin{thebibliography}{99}

\bibitem{geometricphases} A. Shapere and F. Wilczek, \textit{Geometric Phases in Physics} (World Scientific, Singapore, 1989).

\bibitem{berry} M. V. Berry, Proc. R. Soc. London A {\bf 392}, 45 (1984).

\bibitem{ab} Y. Aharonov and D. Bohm, Phys. Rev. {\bf 115}, 485 (1959).

\bibitem{ac} Y. Aharonov and A. Casher, Phys. Rev. Lett. {\bf 53}, 319 (1984).

\bibitem{ra} B. Reznik and Y. Aharonov, Phys. Rev. D {\bf 40}, 4178 (1989).

\bibitem{bjvanwees} B. J. van Wees, Phys. Rev. Lett. {\bf 65}, 255 (1990).

\bibitem{esimanek} E. Simanek, Phys. Rev. B {\bf 55}, 2772 (1997).

\bibitem{ivanov} D. A. Ivanov, L.\ B.\ Ioffe, V.\ B.\ Geshkenbein, and G.\ Blatter,
Phys. Rev. B. {\bf 65}, 024509 (2001).

\bibitem{friedman} J. R. Friedman and D. V. Averin, Phys. Rev. Lett. {\bf 88}, 50403 (2002).

\bibitem{ldvg} D. Loss, D. P. DiVincenzo and G. Grinstein, Phys. Rev. Lett. {\bf 69}, 3232 (1992).

\bibitem{ws} W. Wernsdorfer and R. Sessoli, Science {\bf 284}, 133 (1999).


\bibitem{oetal} T.\ P.\ Orlando, J.\ E.\ Mooij, L.\ Tian, C.\ H.\ van der Wal, 
L.\ S.\ Levitov, S.\ Lloyd, and J.\ J.\ Mazo, Phys. Rev. {\bf B60},
15398 (1999).


\bibitem{blatter} G. Blatter, V. B. Geshkenbein and L. B. Ioffe, Phys. Rev. B {\bf 63}, 174511 (2001).





\bibitem{mooij} C.\ H.\ van der Wal and J.\ E.\ Mooij, J.\ Supercond. {\bf 12}, 807 (1999). 








\end{thebibliography}
\end{document}